\documentclass[preprint,12pt]{elsarticle}

\usepackage{amsmath}
\usepackage{bm}
\usepackage{xspace}

\newcommand{\sNN}{\ensuremath{s_\mathrm{NN}}\xspace}

\usepackage{amsmath}
\usepackage{mathtools}
\usepackage{mathrsfs}
\usepackage{amssymb}
\usepackage{dsfont}
\usepackage{slashed}
\usepackage{empheq}
\usepackage{relsize}
\usepackage{nicefrac}
\usepackage{enumerate}

\usepackage{braket}
\usepackage{amsmath}
\usepackage{mathtools}
\usepackage{mathrsfs}
\usepackage{amssymb}
\usepackage{dsfont}
\usepackage{slashed}
\usepackage{empheq}
\usepackage{relsize}
\usepackage{nicefrac}
\usepackage{amsmath}
\usepackage{bm}
\usepackage{multirow}

\usepackage[utf8]{inputenc}
\usepackage[ngerman,english]{babel}
\usepackage{mathtools}

\usepackage{bm}
\usepackage{epstopdf}
\usepackage{amsbsy}
\usepackage{amsmath}
\usepackage{amssymb}
\usepackage{siunitx}
\usepackage{commath}
\usepackage{graphicx}
\usepackage{graphics} 

\newcommand{\EQ}[3]{
  \begin{equation}
    \label{#1}
    #2
    \;#3
  \end{equation}
}
\newcommand{\ER}[1]{Eq.~(\ref{#1})}
\newcommand*\diff{\mathop{}\!\mathrm{d}}
\newcommand*\Diff[1]{\mathop{}\!\mathrm{d^#1}}

\begin{document}

\begin{frontmatter}


\title{Limiting fragmentation at LHC energies}


\author{B.~Kellers}
\author{G.~Wolschin\corref{cor}}
\ead{g.wolschin@thphys.uni-heidelberg.de}

\address{Institut f{\"ur} Theoretische Physik der Universit{\"a}t Heidelberg, Philosophenweg 12-16, D-69120 Heidelberg, Germany, EU}

\cortext[cor]{Corresponding author}
\begin{abstract}
We investigate the validity of the limiting-fragmentation hypothesis in relativistic heavy-ion collisions at energies reached at the Large Hadron Collider (LHC).
A phenomenological analysis of central AuAu and PbPb collisions 
based on a three-source relativistic diffusion model (RDM) is used to extrapolate pseudorapidity distributions of produced charged hadrons from RHIC to LHC energies into the fragmentation region. Data in this region are not yet available at LHC energies, but our results are
compatible with the limiting-fragmentation conjecture in the full energy range 
$\sqrt{\sNN} = \SI{19.6}{\GeV}\text{ to }\SI{5.02}{\TeV}$.
\end{abstract}

\begin{keyword}
Relativistic heavy-ion collisions \sep LHC energies \sep Limiting fragmentation \sep Produced charged hadrons 

\PACS{25.75.-q \sep24.10.Jv \sep 24.60.-k }

\end{keyword}

\end{frontmatter}
\newpage
\section{Introduction}
The significance of the fragmentation region in relativistic heavy-ion collisions was realized when data on AuAu collisions in the energy range $\sqrt{\sNN}$ = \SIrange{19.6}{200}{\GeV} became available at the Brookhaven Relativistic Heavy Ion Collider (RHIC) \cite{bea02,bb03,ada06}. For a given centrality, the pseudorapidity distributions 
of produced charged particles were found to scale with energy according to the limiting fragmentation (LF), or extended longitudinal scaling, hypothesis: 
The charged-particle pseudorapidity distribution is energy independent over a large range of pseudorapidities $\eta\,'=\eta-y_\text{beam}$, with the beam rapidity $y_\text{beam}$.

The existence of the phenomenon had been predicted for hadron-hadron and electron-proton collisions by Benecke et al. \cite{benecke69}, and it was first shown to be present in 
$p\bar{p}$ data, in a range from \num{53} up to \SI{900}{\GeV} \cite{al86}. 
The fragmentation region grows in pseudorapidity with increasing collision energy and can cover more than half of the pseudorapidity range over which particle production occurs. The approach to a universal limiting curve is a remarkable feature of the particle production process, especially in relativistic heavy-ion collisions.

Presently it is not clear, however, whether limiting fragmentation will persist at the much higher incident energies that are available at the CERN Large Hadron Collider (LHC), namely,
$\sqrt{s_\text{NN}}$ = 2.76 and 5.02 TeV in PbPb collisions. Although detailed and precise ALICE data for charged-hadron production at various centralities are available in the midrapidity region \cite{alice276,alice502} for both incident energies, experimental results in the fragmentation region are not available due to the lack of a dedicated forward spectrometer. This region is most interesting if one wants to account for the collision dynamics more completely. In this work, we investigate to what extent limiting fragmentation can be expected to occur in heavy-ion collisions at LHC energies.

Given the lack of LHC data in the fragmentation regions, one has to rely on either microscopic approaches such as the multiphase transport model AMPT by Ko et al.\;\cite{ko05} or HIJING \cite{wang91,papp18} in order to assess whether LF is valid at LHC energies, or on phenomenological models. Among these are the thermal model \cite{hag65,pbm95,mabe08,pbm16}, hydrodynamical approaches \cite{hesne13}, or the relativistic diffusion model (RDM) with three sources for particle production: a midrapidity source and two fragmentation sources \cite{wolschin99,biya02,wobi06,wob06,wolschin13,gw16}.
Here, the time evolution of the distribution functions is accounted for through solutions of a Fokker-Planck equation (FPE) for the rapidity variable which are subsequently transformed to pseudorapidity space through the appropriate Jacobian.

Regarding microscopic approaches, the AMPT code \cite{ko05} had been tuned for the most central bin at RHIC and LHC energies in Ref.\,\cite{basu16}. There are some disagreements with the LHC data in the midrapidity region, but AMPT is in accord with longitudinal scaling at LHC energies. In Ref.\,\cite{na11} it had already been concluded that AMPT and other microscopic codes reproduce LF at RHIC energies. The same conclusion had been drawn from calculations in the color-glass-condensate framework \cite{sta06}.

The ALICE collaboration has argued in Ref.\,\cite{abb13}, in accordance with our results in Ref.\,\cite{wolschin12}, that their \SI{2.76}{\TeV} PbPb data are in agreement with the validity of extended longitudinal scaling -- within the uncertainties which arise mainly from the extrapolation of the charged-particle pseudorapidity density from the measured region to the rapidity region of the projectile where no data are available. In their analysis, the extrapolation into the forward $\eta$-region was done using the difference of two gaussian functions as detailed in Ref.\,\cite{abb13}. Both Gaussians are centered at midrapidity, and the second (subtracted) Gaussian simulates the central dip that is mostly due to the jacobian transformation from $y$- to $\eta$-space. The same extrapolation function has recently been used in Ref.\,\cite{sahoo19} at both LHC energies.

This procedure leads ALICE to conclude that the 2.76 TeV PbPb data are consistent with the LF hypothesis. Related, but different, extrapolation schemes give similar conclusions.
As an example, we have fitted the \mbox{ALICE} midrapidity data with a sum of two Gaussians that are peaked at the experimental maxima,
used the proper jacobian transformation from $y$- to $\eta$-space, and determined the corresponding parameters for PbPb collisions in a $\chi^2$-minimisation. Again, the resulting pseudorapidity distribution functions fulfill the LF hypothesis at both LHC energies, 2.76 and 5.02 TeV. However, such an extrapolation procedure is a rather arbitrary scheme without any physical basis. In contrast, the thermal model \cite{hag65,pbm95,mabe08,pbm16} has a macroscopic physical basis, which is appropriate to predict particle production rates at midrapidity -- but it is questionable whether it is suited to predict distribution functions, in particular at forward rapidity. Still, it has been used in Ref.\,\cite{cleymans08} in the forward region, with the conclusion that limiting fragmentation should be violated at LHC energies.

In this work, we investigate whether limiting fragmentation in heavy-ion collisions at LHC energies can be expected to be fulfilled in yet another phenomenological model, the three-source relativistic diffusion model (RDM) \cite{wolschin99,wolschin13,gw16}. We briefly summarise the basic formulation in the next section. In section 3, we apply the model both in its analytically solvable version with linear drift, and with a sinh-drift term that requires a numerical solution, to calculate net-proton rapidity distributions at RHIC energies. Here, only the fragmentation sources contribute, and thus can be directly compared to data. In section 4 we apply the model, including a central source, to charged-hadron production in central
AuAu and PbPb collisions at RHIC and LHC energies, in order to test whether limiting fragmentation is fulfilled at LHC energies. The conclusions are drawn in section 5.

\section{A phenomenological three-source model}
In relativistic heavy-ion collisions, the relevant observable in stopping and particle production is the Lorentz-invariant cross section
\EQ{eq:3e}{E \frac{\Diff3 N}{\diff p^3} = \frac{\Diff2 N}{2 \pi p_\perp \diff p_\perp \diff y} = \frac{\Diff2 N}{2 \pi m_\perp \diff m_\perp \diff y}}{}
with the energy $E = m_\perp\cosh(y)$, the transverse momentum $p_\perp = \sqrt{p_{x}^2+p_{y}^2}$, the transverse mass $m_\perp = \sqrt{m^2+p_\perp^2}$,  and 
the rapidity $y$. 

\begin{figure}[h]
\begin{center}
\includegraphics[scale=0.8]{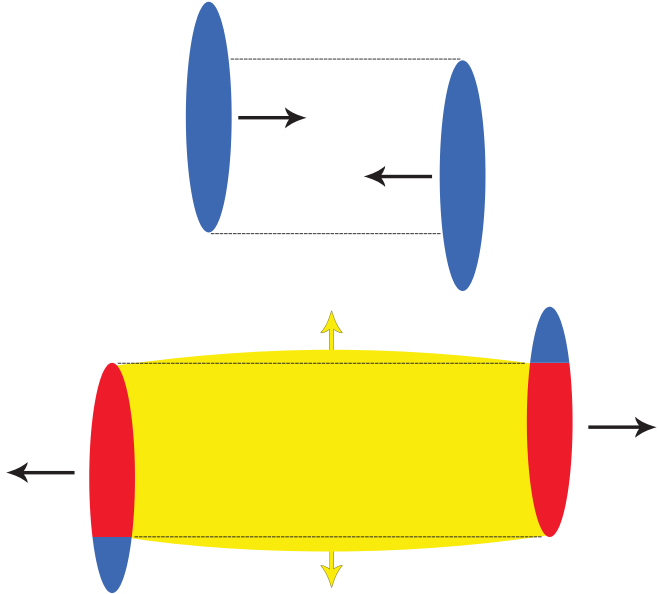}
\caption{ Schematic representation of the three-source model for particle production in relativistic heavy-ion collisions at RHIC and LHC energies in the center-of-mass system: Following the collision of the two Lorentz-contracted slabs (blue), the fireball region (center, yellow) expands anisotropically in longitudinal and transverse direction. At midrapidity, it represents the main source of particle production. The two fragmentation sources (red) contribute to particle production, albeit mostly in the forward and backward rapidity regions.}
\label{fig1}
\end{center}
\end{figure}

We shall first investigate rapidity distributions of protons minus produced antiprotons, which are indicative of the stopping process as described 
phenomenologically in a relativistic two-source diffusion model (RDM) {\cite{wolschin99,wolschin17}} or in a QCD-based approach \cite{mtw09}. This motivates the relevance of the fragmentation sources not only in stopping, but also in particle production at relativistic energies. 
Subsequently, we switch to a three-source model for particle production, with the importance of the fireball source rising with energy, and contributing most of the produced charged hadrons at LHC energies when compared to the fragmentation sources. This central source does not contribute to stopping because particles and antiparticles are produced in equal amounts. The three-source model is visualized schematically in Fig.\,\ref{fig1} for a symmetric system such as AuAu or PbPb.

The rapidity distributions for all three sources $k=1,2,3$ are obtained by integrating over the transverse mass 
\EQ{rap}{
  \frac{\diff N_k}{\diff y}(y,t) = c_{\,k} \int\; m_\perp E \frac{\Diff3 N_k}{\diff p^3} \diff m_\perp
}{,}
with normalisation constants $c_k$ that depend on centrality. The experimentally observable distribution {$\text{d}N/\text{d}y$ is evaluated in the time-dependent model at the freeze-out time, $t = \tau_\text{\,f}$\,. The latter can be identified with the interaction time $\tau_\text{int}$ of Refs.\,\cite{wolschin99,wolschin17}: the time during which the system interacts strongly. The full rapidity distribution function for produced charged hadrons is obtained by weighting the three partial distribution functions with the respective numbers of particles, and adding them incoherently:
\begin{equation}
\frac{\text{d}N_\text{ch}}{\text{d}y}(y,t=\tau_\text{f})=N_\text{ch}^{1}R_{1}(y,\tau_\text{f})
 +N_\text{ch}^{2}R_{2}(y,\tau_\text{f})
+N_\text{ch}^\text{gg}R_{\text{gg}}(y,\tau_\text{f})\,,
\label{normloc1}
\end{equation}
where the index ${3\equiv\text{gg}}$ is meant to emphasise that the fireball source is mostly arising microscopically from low-$x$ gluon-gluon collisions.

The incoherent addition of the three sources applies also to the model with sinh-drift that we consider in this work, because the FPE is a linear partial differential equation, allowing for linear superposition of independent solutions. For a symmetric system, one can further simplify the problem by only considering the solution for the positive rapidity region and mirroring the result at $y<0$.

The parameters of the three-source model -- which will be detailed in the following -- are then determined via $\chi^2$-minimisation with respect to the available data, and can be used in extrapolations and predictions \cite{gw16}. In stopping, the relevant distribution function is given by the incoherent sum of the fragmentation sources only,
\begin{equation}
\frac{\text{d}N_{p-\bar{p}}}{\text{d}y}(y,t=\tau_\text{f})=N_{p-\bar{p}}^{1}R_{1}(y,\tau_\text{f})
 +N_{p-\bar{p}}^{2}R_{2}(y,\tau_\text{f})\,.
\label{normloc2}
\end{equation}
We rely on Boltzmann-Gibbs statistics and hence adopt  the Maxwell-J{\"u}ttner distribution
as the thermodynamic equilibrium distribution for $t\rightarrow \infty$ 
\begin{equation}
\label{equ}
  E \frac{\text{d}^3N}{\text{d}p^3} \Bigr|_\text{eq}\propto E \exp\left(-E/T\right)
  = m_\perp \cosh\left(y\right) \exp\left(-m_\perp \cosh(y) / T\right).
\end{equation}
The nonequilibrium evolution of all three partial distribution functions $R_k(y,t)$ $(k=1,2,gg)$ towards this thermodynamic equilibrium distribution is accounted for in the relativistic diffusion model \cite{wolschin99,biya02,wolschin13,wolschin17}  through solutions of the Fokker-Planck equation 
\EQ{fpe}{
  \pd{}{t}R_k(y,t) = -\pd{}{y}\left[J_k(y,t)R_k(y,t)\right] + \pd[2]{}{y}\left[D_k(y,t)R_k(y,t)\right]
}\\
with suitably chosen drift functions $J_k(y,t)$ and diffusion functions $D_k(y,t)$. If the latter is taken as a constant diffusion coefficient $D_k$,
and the drift function assumed to be linearly dependent on the rapidity variable $y$, 
the FPE has the Ornstein-Uhlenbeck form \cite{uo} and can be solved analytically \cite{wolschin99}. For $t\rightarrow \infty$ all three subdistributions approach a single Gaussian in rapidity space which is centered at midrapidity $y=0$ for symmetric systems, or at the appropriate equilibrium value $y=y_\text{eq}$ for asymmetric systems. In case of stopping, only the two fragmentation distributions contribute,  approaching the thermal equilibrium distribution for $t\rightarrow \infty$, as will be shown in the next section.

It should be noted that interpenetration and stopping (or more precisely, slowing down) of the lorentz-contracted, highly transparent nuclei occurs before the QGP-medium with quarks and gluons in the fireball is fully formed. Hence, there exists no medium or heat bath that could act as a solvent providing friction and noise due to thermal fluctuations, as is the case in the diffusion model for Brownian motion, or for heavy quarks in a QGP. Instead, the incident baryons loose their momentum (rapidity) without any globally static medium, but through random partonic two-body collisions between valence quarks and low-$x$ gluons in the respective other nucleus. These provide the fluctuating environment necessary for the formulation of a Langevin equation, or equivalenty, the corresponding Fokker--Planck equation for the relativistic system.

The FPE Eq.\,(\ref{fpe}) in the context of relativistic heavy-ion collisions can be derived from a theory for 
non-Markovian processes in spacetime, which are equivalent to relativistic Markov processes in phasespace
(RMPP), see Refs.\,\cite{denisov09,dunkel09}. It is shown in these works that such markovian processes in phasespace are accounted for through a generalised FPE. 
The basic equation Eq.\,(\ref{fpe}) in rapidity space that we are using in the present work is a special case in the context of such a more general RMPP formalism.  

The equilibrium limit of the FPE solution for constant diffusion and linear drift is, however, found to deviate slightly from the Maxwell-J{\"u}ttner distribution. Although the discrepancies are small and become visible only for sufficiently large times, we use the RDM with the sinh-drift
\begin{equation}
J_k(y,t)=-A_k \sinh(y)\,,
\end{equation}
which ensures that the solution for $t\rightarrow \infty$ yields the Maxwell-J{\"u}ttner distribution Eq.\,(\ref{equ}), as was discussed in Refs.\,\cite{lavagno02,wolschin18}. 
This induces a special form of the fluctuation-dissipation theorem (FDT) that connects diffusive and dissipative phenomena in the collision, namely
\begin{equation}
A_k= m_\perp D_k/T\,.
\label{fdt}
\end{equation}
The strength of the drift force in the fragmentation sources $k = 1,2$ depends on the distance in $y$-space from the beam rapidity, which enters through the initial conditions.
With Eqs.\,(\ref{rap}) and (\ref{equ}), the rapidity distribution at thermal equilibrium can then be derived \cite{wolschin17} as
\begin{equation}
\frac{\text{d}N_\text{eq}}{\text{d}y}=C \left( m_\perp^2 T + \frac{2 m_\perp T^2}{\cosh y} + \frac{2 T^3}{\cosh^2 y} \right) \exp\left({-\frac{m_\perp \cosh y}{T}}\right),
\label{eqfdt}
\end{equation}
where $C$ is proportional to the overall number of produced charged hadrons  $N^\text{tot}_\text{ch}$, or -- in case of stopping --  to the number of net baryons (protons) in the respective centrality bin. Since the actual distribution functions remain far from thermal equilibrium, the total particle number is evaluated based on the nonequilibrium solutions of the FPE, which are adjusted to the data in $\chi^2$-minimisations. In particular,
one can determine the drift amplitudes $A_k$ from the position of the fragmentation peaks as inferred from the data, and then calculate theoretical diffusion coefficients as $D_k=A_k T/m_\perp$. These refer, however, only to the diffusive processes, and since the fireball source and both fragmentation sources also expand collectively, the actual distribution functions will be much broader than what is obtained from Eq.\,(\ref{fdt}). Hence, we shall use values
for the diffusion coefficients (or the widths of the partial distributions) that are adapted to the data in both stopping and particle production. The total particle number is then obtained from the integral of the overall distribution function.

Whereas the RDM with linear drift has analytical solutions that can be used directly in $\chi^2$-minimisations with respect to the data, numerical solutions of the FPE are required for the sinh-drift, as described in Refs.\,\cite{wolschin17,wolschin18}.
To arrive at a usable form for the computer{,} we transform the equation for $R(y,t)$ into its dimensionless version for $f(y,\tau)$ by introducing a timescale $t_c$, defining the dimensionless time variable $\tau = t / t_c$.
It follows that $\pd{}{t} = \pd{}{\tau} t_c^{-1}$ and hence
\EQ{eq:4}{
  \pd{f}{\tau}(y,\tau) = t_c\;A\;\pd{}{y}\left[\sinh(y)\;f(y,\tau)\right] + t_c\;D\;\pd[2]{}{y}f(y,\tau)
}{.}
Since {$A = m_\perp D/T$}, we set {$t_c =T/(m_\perp D) = A^{-1}$}.
The result is the dimensionless \ER{eq:4aa} depending only on the ratio {$\gamma = T/m_\perp$} of temperature $T$ and transverse mass $m_\perp$ which is a measure of the strength of the diffusion,
\EQ{eq:4aa}{
    \pd{f}{\tau}(y,\tau) = \pd{}{y}\left[\sinh(y)\;f(y,\tau)\right] + \gamma\;\pd[2]{}{y}f(y,\tau)
}
{.}
To recover the drift and diffusion coefficients, one has to specify a time scale (or the other way round).
Considering that it is only the drift term that is responsible for determining the peak position{,} we choose the time-like variable $\tau$ such that the peak position of the experimental data is reproduced.
This leaves the diffusion strength $\gamma$ as free parameter. In case of three partial distributions, there are three free parameters $\gamma_k$. Here, the two values for the fragmentation sources are identical for symmetric systems such as PbPb, but differ for asymmetric systems like $p$Pb.

We calculate the numerical  solution using \textsc{matlab}'s integration routine \texttt{pdepe} for solving parabolic-elliptic partial differential equations.
It was shown in Ref.\,\cite{wolschin17} that this method is very accurate when compared to results of finite-element methods such as DUNE \cite{ba10} and FEniCS \cite{aln15}. 

To compare the simulation to experimental data, we have to insert relevant values for $T$, $m_\perp$, and the initial conditions.
The beam rapidity $y_\text{beam}$ is determined by the center-of-mass energy per nucleon pair as $y_\text{beam} =\pm \ln(\sqrt{s_{NN}}/m_p)$.
Two gaussian distributions centered at the beam rapidities with a small width that corresponds to the Fermi motion represent the incoming ions before the collision.
The exact width of the initial distribution does not have a large effect on the time evolution \cite{wolschin17}; here we use $\sigma=0.1$. 
The same standard deviation is taken for the initial condition of
the midrapidity source, which is centered at $y=0$ for a symmetric system, and at  $y=y_\text{eq}$ for asymmetric systems. 

For the temperature{,} we take the critical value $T=T_\text{cr}=\SI{160}{\MeV}$ for the cross-over transition between hadronic matter and quark-gluon plasma. Regarding the transverse mass,
experimental values are deduced from measured transverse-momentum distributions.

 
 The results are then transformed to rapidity distributions \cite{wolschin17}.
 Rewriting Eq.\,(\ref{rap}) and replacing {$\text{d}^3N/\text{d}p^3$} with the computed
distribution $f(y,\tau)$, we obtain
\EQ{rapf}
{
    \od{N}{y}(y,\tau) = C \int m^2_\perp f(y,\tau) \dif m_\perp \\\,.
}{}
The constant ${C}$ is chosen in case of stopping such that the total number of particles corresponds to the number of participant protons in the respective centrality bin. In particle production, $C$ is adjusted to the total number of produced particles for a given centrality. 
\section{Fragmentation sources in stopping}
To emphasise the relevance of the fragmentation sources, we first investigate stopping and calculate net-proton rapidity distributions in central AuAu collisions at RHIC energies of \SI{200}{\GeV}, where data are available from Ref.\,\cite{bea04}.
Theoretical calculations are usually performed for net-baryon distributions \cite{mtw09} because the total baryon number is a conserved quantity,
but since experimentally only net-proton distributions are available, we convert to net-protons via $Z/A=79/197=0.40$.
As discussed, the central source cancels out in stopping, because particles and antiparticles are produced in equal amounts.
\begin{figure}[t!]
\begin{center}
\includegraphics[scale=0.6]{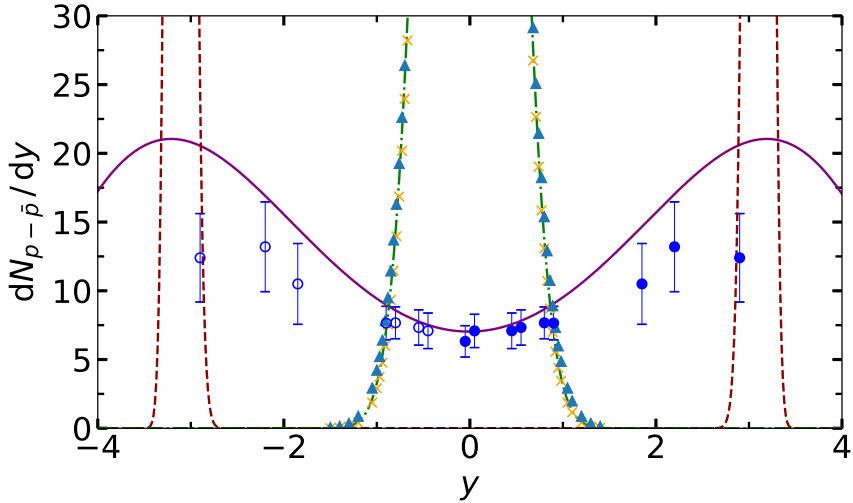}
\caption{ The fragmentation sources are visible in net-proton (proton minus antiproton, or stopping) rapidity density functions for central AuAu collisions at $\sqrt{s_\mathrm{NN}}=\SI{200}{\GeV}$. The BRAHMS data at forward rapidities (filled circles from Ref.\,\cite{bea04}) have been symmetrized (open circles at backward rapidities). Solid curves correspond to the numerical model with sinh-drift for $\gamma \equiv \gamma_{1,2} = 33$ and $\tau = 0.08$, with stopping peaks at $y=\pm\, 3.1$. The dashed curves are also calculated in the sinh-drift model, but with $\gamma = 0.139$ as predicted by the fluctuation-dissipation relation Eq.\,(\ref{fdt}), ignoring collective expansion. The dot-dashed curve in the midrapidity region represents the equilibrium limit in the sinh-model with $\tau \rightarrow \infty$ and $\gamma = 0.139$. It agrees with the Maxwell-J{\"u}ttner distribution for $m_\perp = \SI{1.15}{\GeV}$ and $T = \SI{160}{\MeV}$ (crosses).  The equilibrium limit of the analytical linear-drift model for $t \rightarrow \infty$ (triangles) deviates only slightly from the Maxwell-J{\"u}ttner distribution. The fireball source does not contribute to stopping.} 
\label{fig2}
\end{center}
\end{figure}
Rather precise net-proton results had been obtained at SPS energies for central PbPb at $\sqrt{s_\text{NN}}=\SI{17.3}{\GeV}$,
where the NA49 collaboration succeeded to measure across the fragmentation peak in a fixed-target experiment \cite{app99}. These results can be well
reproduced in the linear RDM \cite{gw04}, and also in a QCD-based approach \cite{mtw09}, but here we are interested in higher energies, namely, the RHIC and LHC region.
\begin{figure}[h]
\begin{center}
\includegraphics[scale=0.6]{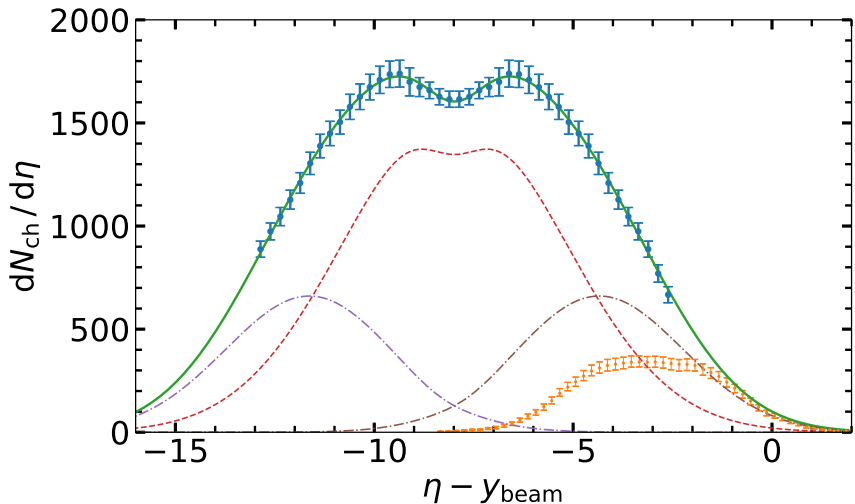}
\caption{ Pseudorapidity density distribution in the three-source RDM with linear drift (solid curve, top) resulting from a $\chi^2$-minimisation with respect to the ALICE data \cite{alice276} for produced charged hadrons in central (0-5\%) PbPb collisions at $\sqrt{s_\mathrm{NN}}=\SI{2.76}{TeV}$. The distributions are displayed as functions of $\eta-y_\text{beam}$. Central AuAu collision data at $\sqrt{s_\mathrm{NN}}=\SI{19.6}{\GeV}$ (lower data points, \cite{alver11}) are consistent with the limiting-fragmentation hypothesis (LF). The total density distribution can be separated into three parts, one resulting from the midrapidity source (dashed curve) and two from the fragmentation sources (dot-dashed curves).} 
\label{fig3}
\end{center}
\end{figure}
For AuAu at a RHIC energy of \SI{200}{\GeV}, it was not possible to measure the fragmentation peaks, because the forward spectrometer hit the beam pipe at large rapidities. At LHC energies, a forward spectrometer with particle identification in the region of the expected fragmentation peaks \cite{mtw09} is not available. Still, the BRAHMS stopping data of Ref.\,\cite{bea04} shown in Fig.\,\ref{fig2} indicate the rise towards the fragmentation peaks, which was later corroborated by more recent -- albeit preliminary -- data near the peak region \cite{deb08}.

We now compare these AuAu RHIC data with the fragmentation distributions that arise from the three-source model with both linear and sinh-drift.
In case of a linear drift, the average positions of the fragmentation peaks agree with the maximum-value positions, $y_\mathrm{peak} = \langle y_{1,2} \rangle$, whereas these differ when the drift is nonlinear.  
For AuAu the transverse mass $m_\perp$ is obtained from the $p_\perp$-spectra using $m_\perp=\sqrt{m_p^2+p_\perp^2}$ and Eq.\,(\ref{eq:3e}) for the yields. 
As proposed in Ref.\,\cite{bea04}, Gaussians are fitted to the invariant yields \cite{wolschin17}. The results for protons and antiprotons are averaged to obtain
\begin{equation}
\langle m_\perp\rangle=\SI{1.15+-0.20}{\GeV},
\label{mt}
\end{equation}
which we use in the forthcoming stopping calculations. The theoretical value for the diffusion strength becomes $\gamma=T/m_\perp=0.139$. 

The results of the RDM-calculation with the sinh-drift are shown as dashed curves in Fig.\,\ref{fig2}. The dimensionless time parameter has been adjusted as $\tau=0.08$ with the
above value of the diffusion strenght $\gamma$ to yield a fragmentation-peak position of $y=\pm 3.1$, in accordance with the data of Refs.\,\cite{bea04,deb08} (only the final data of 
Ref.\,\cite{bea04} are shown here). The calculated distribution function is, however, by far too narrow, because the theoretical expression from Eq.\,(\ref{fdt}) does not account for collective expansion. In Ref.\,\cite{gw06}, the longitudinal expansion velocity $v_{||}$ had actually been calculated from the difference between the theoretical distribution function, and the data. The solid curve represents a numerical solution of the FPE with adapted diffusion strength $\gamma=33$, it clearly shows the fragmentation peaks.

In Fig.\,\ref{fig2} we also display the corresponding equilibrium solutions, which are centered at midrapidity for symmetric systems. We use the theoretical FDT-value $\gamma=0.139$, and display the numerical solution of the FPE with sinh-drift for $\tau \rightarrow \infty$ as dot-dashed curve. It agrees with the Maxwell-J{\"u}ttner distribution Eq.\,(\ref{equ}) for $m_\perp= \SI{1.15}{\GeV}$ and $T = \SI{160}{\MeV}$ (crosses). For comparison, the equilibrium result of the RDM with linear drift is also shown (triangles). It is a Gaussian
\begin{equation}
R_\text{eq}(y)={\frac{C}{\sqrt{2\pi\sigma_\text{eq}}} \exp\bigg(-\frac{y^2}{2\sigma^2_\text{eq}}}\bigg)
\label{gauss}
\end{equation}
 with a variance $\sigma_\text{eq}^2=\gamma=t_cD=A^{-1}D=T/m_\perp=0.139$ corresponding to a width $\Gamma_\text{FWHM}=\sqrt{8\gamma\ln2}=0.88$. The normalisation $C$ is such that the integral of the total distribution yields the number of participant protons in $0-5\%$ central AuAu collisions, $N_p=N_\text{B}Z/A=357\times79/197\simeq 143$, with the number of participant baryons $N_\text{B}=357\pm8$ from a Glauber calculation \cite{bea04}. The resulting distribution function deviates only slightly from the Maxwell-J{\"u}ttner equilibrium distribution. 

Equilibrium distributions with physical values for the diffusion strengths that include collective expansion would be much broader, but they do not exhibit fragmentation peaks with a midrapidity valley, and hence, equilibrium models are not suited to describe stopping distributions.

\section{Particle production and limiting fragmentation}
In charged-hadron production, we consider the sum of produced charged particles and antiparticles. Hence, the fireball source has to be added, and yields the essential contribution to charged-hadron production in heavy-ion collisions at LHC energies. Particles that are produced from the fragmentation sources are not directly distinguishable from those originating from the fireball, but still the fragmentation sources are relevant and must be included in a phenomenological model. In particular, when regarding the limiting-fragmentation conjecture, the role of the fragmentation distributions will turn out to be decisive since they determine the behavior of the rapidity distribution functions at large values of rapidity. 

For unidentified charged particles, we first have to transform from rapidity- to pseudorapidity space in order to directly compare to data. The pseudorapidity variable $\eta$ is uniquely determined by the scattering angle $\theta$ 
\begin{equation}
\eta=\frac{1}{2}\ln\frac{|\bf{p}|+\rm p_{||}}{|\bf{p}|-\rm p_{||}}=-\ln\left[\tan\left(\theta/2\right)\right]\,,
\label{eta}
\end{equation}
and the pseudorapidity distribution function $\frac{\mathrm{d}N}{\mathrm{d}\eta}$ is obtained from the rapidity distribution $\frac{\mathrm{d}N}{\mathrm{d}y}$ through the transformation
\begin{align}
	\frac{\mathrm{d}N}{\mathrm{d}\eta} = \frac{\mathrm{d}y}{\mathrm{d}\eta} \frac{\mathrm{d}N}{\mathrm{d}y} = \mathrm{J}\left(\eta, \frac{m}{p_\perp}\right) \,\frac{\mathrm{d}N}{\mathrm{d}y}\,,
	\label{dNdeta}
\end{align}
with the Jacobian
\begin{align}
\mathrm{J} \left(\eta, \frac{m}{p_\perp}\right) = 
		\frac{\cosh(\eta) }{\sqrt{ 1 + \left(\frac{m}{p_\perp}\right)^2 + \sinh^2(\eta) }}
	\label{jac}
\end{align}
for produced particles with mass $m$ and transverse momentum $p_\perp$.
The transformation depends on the squared ratio $(m$/$p_\perp)^2$ of mass and transverse momentum of the produced particles. Hence, its effect increases with the mass of the particles and is most pronounced at small transverse momenta. In principle, one has to consider the full $p_\perp$-distributions, which are, however, not available for all particle species that are included in the pseudorapidity measurements. 
In Ref.\,\cite{wolschin12}, we have determined the Jacobian $\mathrm{J}_0$ at $\eta=y=0$ in central 2.76 TeV PbPb collisions for identified $\pi^-, K^-$, and antiprotons from the experimental values $\frac{\mathrm{d}N}{\mathrm{d}\eta}|_\text{exp}$ and $\frac{\mathrm{d}N}{\mathrm{d}y}|_\text{exp}$ as $\mathrm{J}_0=0.856$. Solving Eq.\,(\ref{jac}) for $p_\perp\equiv \langle p_\perp^{\text{eff}}\rangle$ yields 
\begin{equation}
\langle p_\perp^{\text{eff}}\rangle=\frac{\langle m \rangle \mathrm{J}_0}{\sqrt{1-\mathrm{J}_0^2}}
\label{pteff}
\end{equation}
with a mean mass $\langle m \rangle$  which may be calculated from the abundancies of pions, protons and kaons. With the introduction of $\mathrm{J}_0$,
the Jacobian can then be written independently from the values of $\langle m \rangle$ and $\langle p_\perp^{\text{eff}}\rangle$ as
\begin{equation}
\mathrm{J} \left(\eta, \mathrm{J}_0\right) = 
	\frac{\cosh(\eta) }{\sqrt{ 1 +\frac{1-\mathrm{J}_0^2}{\mathrm{J}_0^2}+ \sinh^2(\eta) }}\,,
	\label{jac1}
\end{equation}
which results in $\mathrm{J}(\eta)=\cosh(\eta)[{1.365+\sinh^2(\eta)}]^{-1/2}$ for central 2.76 TeV PbPb collisions.
The effect of the Jacobian is most pronounced near midrapidity, where it is essential to generate the dip in the pseudorapidity distributions, as is obvious from Fig.\,\ref{fig3}: A calculation in the RDM with linear drift \cite{gw16} is compared with ALICE data for central PbPb at 2.76 TeV \cite{alice276}, with five parameters and $\chi^2$-values from Tab.\,\ref{tab1}. The optimization is done using Python.
\begin{table}
\begin{center}
\caption{Parameters in the RDM with linear drift for central (0-5\%) PbPb collisions at 2.76 TeV and 5.02 TeV: Particle content $N_{1,2}$ and $N_\text{gg}$ of the fragmentation and fireball sources, mean rapidities $\langle y_{1,2} \rangle$ of the fragmentation sources, widths $\Gamma_{1,2,\text{gg}}$, $\chi^2$- and $\chi^2/\text{ndf}$-values.}
\begin{tabular}{ccccccccc}\\
\hline 
$\sqrt{s_\mathrm{NN}}$ (TeV) &$y_\text{\,beam}$& $N_{1,2}$ & $N_\mathrm{gg}$ & $\langle y_{1,2} \rangle$ & $\Gamma_{1,2}$ & $\Gamma_\mathrm{gg}$ & $\chi^2$ & $\chi^2 / \mathrm{ndf} $ \\ 
\hline 

2.76 &$\pm 7.987$& 3505 & 10681 &$\pm 3.64$ & 4.98 & 6.38 & 2.44 & 0.07 \\ 

5.02 &$\pm 8.586$& 4113 & 14326 & $\pm 4.67$ & 4.99 & 6.38 & 1.17 & 0.04 \\ 
\hline 
\label{tab1}
\end{tabular} 
\end{center}
\end{table}
\begin{figure}[h!]
\begin{center}
\includegraphics[scale=0.6]{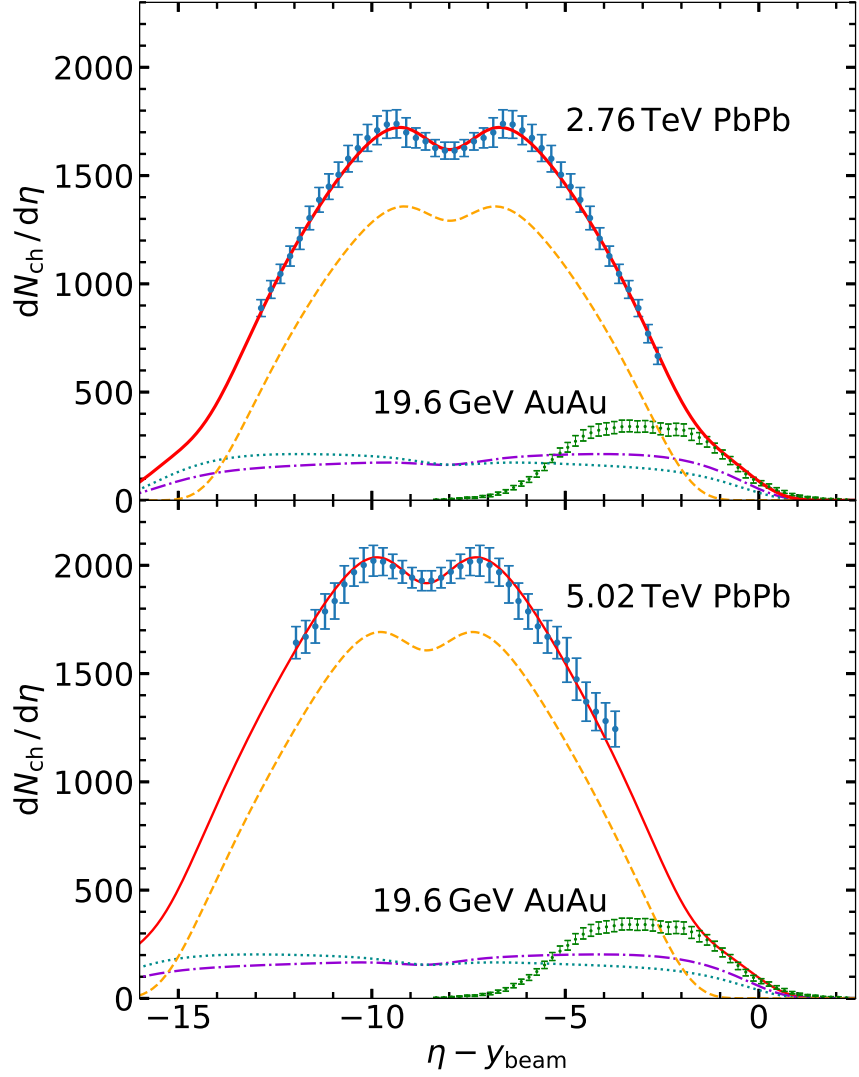}
\caption{ Pseudorapidity density distribution of produced charged hadrons for central (0-5\%) 2.76 TeV (top) and 5.02 TeV PbPb (bottom) in the three-source RDM with sinh-drift (solid curves) from $\chi^2$-minimisations with respect to the ALICE data \cite{alice276,alice502}. The midrapidity sources (dashed curves) remain symmetric, but the fragmentation sources (dotted and dot-dashed curves) are asymmetric due to the sinh-drift. 
Comparison with central \SI{19.6}{\GeV} AuAu RHIC-data \cite{alver11} confirms the consistency with the limiting-fragmentation hypothesis in both cases.} 
\label{fig4}
\end{center}
\end{figure}
\begin{figure}[h!]
\begin{center}
\includegraphics[scale=0.72]{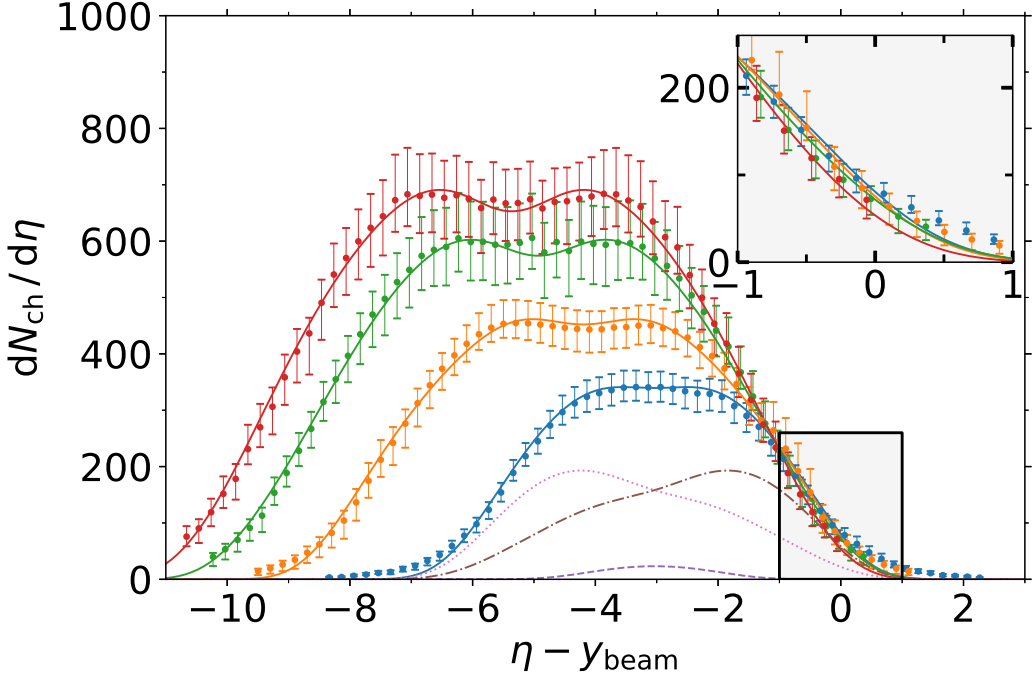}
\caption{ Three-source RDM-distributions with sinh-drift compared to central (0-3\%) PHOBOS AuAu data \cite{alver11} at four RHIC energies,
 $\sqrt{s_\mathrm{NN}}=\,$\,\SIlist{200;130;62.4;19.6}{\GeV} (from top to bottom). The zoom into the fragmentation region shows that the 
 PHOBOS data, and the RDM with sinh-drift are consistent with limiting-fragmentation scaling at RHIC energies. Corresponding model parameters are given in Tab.\,\ref{tab2}. RDM-subdistributions are shown at the lowest energy. } 
\label{fig5}
\end{center}
\end{figure}
\begin{table}
\begin{center}
\caption{Parameters in the RDM with sinh-drift for central (0-3\%) AuAu collisions at 19.6 GeV to 200 GeV: Particle content $N_{1,2}$ and $N_\text{gg}$ of the fragmentation and fireball sources, peak-value rapidities $y_\mathrm{peak}$, mean rapidities $\langle y_{1,2} \rangle$ of the fragmentation sources, diffusion strengths $\gamma_{1,2,\text{gg}}$, corresponding $\chi^2$- and $\chi^2/\text{ndf}$-values.}
\begin{tabular}{cccccccccc}\\
\hline 
$\sqrt{s_\mathrm{NN}}$ (GeV) &$y_\text{\,beam}$& $N_{1,2}$ & $N_\mathrm{gg}$ & $y_\mathrm{peak}$ & $\langle y_{1,2} \rangle$ & $\gamma_{1,2}$ & $\gamma_\mathrm{gg}$ & $\chi^2$ & $\chi^2 / \mathrm{ndf} $ \\ 
\hline 
19.6 &$\pm 3.037$& 870 & 60 &$\pm 0.86$& $\pm 0.42$& 6 & 1 & 157.07 & 3.02 \\

62.4 &$\pm 4.197$& 1280 & 540 & $\pm 1.94$& $\pm 1.11$ & 18 & 4 & 18.11 & 0.37 \\

130 &$\pm 4.931$& 1350 & 1800 &$\pm 2.30$& $\pm 1.08$ & 42 & 13 & 4.07 & 0.08 \\

200 &$\pm 5.362$& 1400 & 2650 & $\pm 2.78$& $\pm 1.64$ & 52 & 24 & 3.39 & 0.07 \\
\hline 
\label{tab2}
\end{tabular} 
\end{center}
\end{table}
\begin{figure}[h!]
\begin{center}
\includegraphics[scale=0.72]{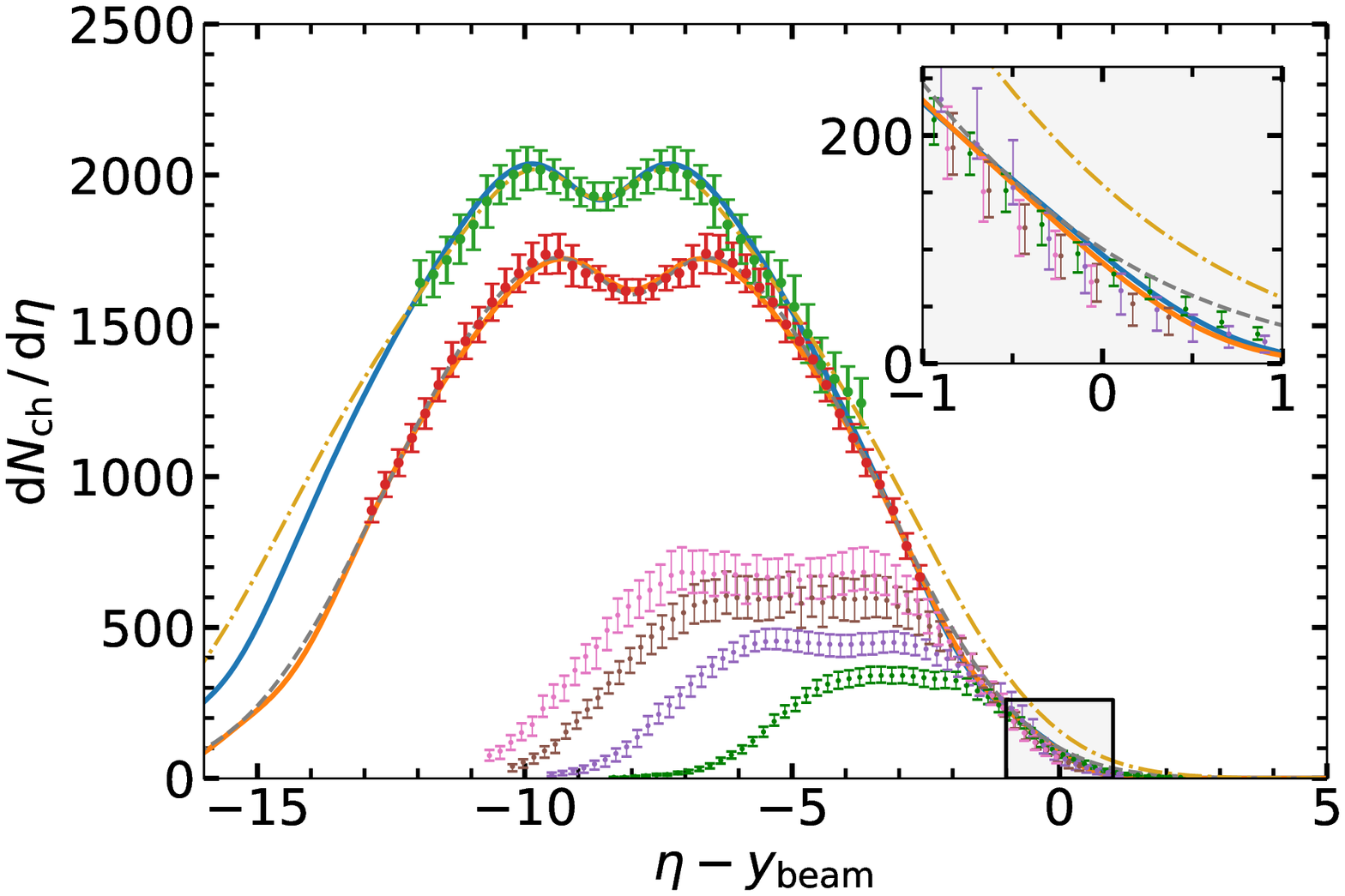}
\caption{ Comparison of the three-source RDM-distributions with linear and sinh-drift, ALICE data \cite{alice276,alice502},  and PHOBOS data \cite{alver11}.
From top to bottom: central PbPb at $\sqrt{s_\mathrm{NN}}=\,$\,\SIlist{5.02;2.76}{TeV} (LHC), AuAu at $\sqrt{s_\mathrm{NN}}=\,$\,\SIlist{200;130;62.4;19.6}{\GeV}. The difference between the model with sinh-drift (solid curves) and the one with linear drift (dot-dashed and dashed curves) is small, but visible in the fragmentation region. The zoom into this region shows that the RDM with sinh-drift is consistent with limiting fragmentation at RHIC and LHC energies.} 
\label{fig6}
\end{center}
\end{figure}

The Jacobian has almost no effect in the fragmentation region, which we are emphasising in this work. In Fig.\,\ref{fig3}, we also compare the RDM-solution for 2.76 TeV PbPb with
central AuAu data \cite{alver11} at $\sqrt{s_\text{NN}} = \SI{19.6}{\GeV}$ from the PHOBOS collaboration at RHIC. In earlier work, we had shown that the three-source RDM-solutions with linear drift agree with PHOBOS data at the RHIC energies of  $\sqrt{s_\text{NN}} = 19.6, 62.4, 130$ and 200 GeV \cite{wob06,kgw07}.

When plotted as function of $\eta-y_\text{beam}$, limiting fragmentation is obviously fulfilled in the relativistic diffusion model with linear drift. This would not be the case if only the central fireball source was considered, as limiting fragmentation is a consequence of the appearance of the fragmentation sources. 
This result indicates that limiting fragmentation can be fulfilled from RHIC to low LHC energies in the relativistic diffusion model.

We now proceed to investigate the consequences of the model with sinh-drift, with emphasis on the fragmentation region. We solve Eq.\,(\ref{eq:4aa}) for central PbPb collisions at 
$\sqrt{s_\text{NN}}=2.76$ and 5.02 TeV using Dirichlet boundary conditions, with parameters given in Tab.\,\ref{tab3}, and the same Jacobian for both energies. The resulting charged-hadron pseudorapidity distributions are shown in Fig.\,\ref{fig4}, again plotted as functions of $\eta-y_\text{beam}$ together with central AuAu data \cite{alver11} at $\sqrt{s_\text{NN}}=\SI{19.6}{\GeV}$ from the PHOBOS collaboration at RHIC. As a consequence of the sinh-drift, the fragmentation distributions are now much less confined to the fragmentation region, but extend into the whole pseudorapidity range that is accessible for produced charged hadrons. Hence, the Jacobian deforms also the fragmentation distributions in the midrapidity region.
In the fragmentation region, LF is very well fulfilled when comparing the results at LHC energies to \SI{19.6}{\GeV} AuAu data.

To confirm that the RDM with sinh-drift is consistent with the observed LF at the available RHIC energies, we compare the PHOBOS AuAu data \cite{alver11}
with the numerical results of our model in Fig.\,\ref{fig5}. At all four energies  $\sqrt{s_\mathrm{NN}}=\,$\,\SIlist{200;130;62.4;19.6}{\GeV} we find
agreement between data and model results, with parameters listed in Tab.\,\ref{tab2}. For the RDM with linear drift, agreement with RHIC data had already been confirmed in our earlier work \cite{wob06}. In both cases, the midrapity source is found to be negligible at 19.6 GeV: The produced-particle yields arise essentially from the fragmentation sources because gluon-gluon collisions are not relevant at this low energy. In the sinh-model, the fragmentation sources are, however, not gaussian in rapidity space, but asymmetric, and extend over a larger rapidity range. This is indicated by the
partial distributions functions shown in Fig.\,\ref{fig5} at this energy. The significance of the midrapidity source rises gradually with increasing energy, see the corresponding particle numbers in Tab.\,\ref{tab2}.

Interestingly, the value for the diffusion strength $\gamma_{1,2} = 52$ in the fragmentation sources for 200 GeV AuAu (see Tab.\,\ref{tab2}) is somewhat larger than the value of  $\gamma_{1,2} = 33$ that we had extracted from the BRAHMS stopping data at this energy in the model with
nonlinear drift, see Fig.\,\ref{fig1}. The difference of the two values 
underlines the fact that the fragmentation peaks in stopping are 
always closer to the beam rapidity than the corresponding peaks 
in particle production. This is expressed by the larger value of
gamma (smaller rapidity relaxation time) in particle production.
The result is physically reasonable, because the fragmentation sources 
in stopping define the mean $y$-position from which lighter hadrons 
are produced at lower rapidity. 

Our overall results from the relativistic diffusion model with linear and sinh-drift are summarised in Fig.\,\ref{fig6}. We compare data from the fragmentation regions in central AuAu collisions at RHIC energies of \SIlist{19.6;62.4;130;200}{\GeV} \cite{alver11} with ALICE data \cite{alice276,alice502} and our results for central PbPb at LHC energies of 2.76 and 5.02 TeV from the relativistic diffusion model with both linear and sinh-drift. The parameters are summarised in Tab.\,\ref{tab1} and Tab.\,\ref{tab3}. As expected, the RDM with sinh-drift (solid curves) gives a better representation of LF
as compared to the analytical linear model (dashed and dot-dashed curves).
The inset shows that the RDM with sinh-drift is in agreement with the limiting-fragmentation conjecture at the available RHIC and LHC energies.
\begin{table}
\begin{center}
\caption{Parameters in the RDM with sinh-drift for central (0-5\%) PbPb collisions at 2.76 TeV and 5.02 TeV: Particle content $N_{1,2}$ and $N_\text{gg}$ of the fragmentation and fireball sources, peak-value rapidities $y_\mathrm{peak}$, mean rapidities  $\langle y_{1,2} \rangle$ of the fragmentation sources, diffusion strengths $\gamma_{1,2,\text{gg}}$, corresponding $\chi^2$- and $\chi^2/\text{ndf}$-values.}
\begin{tabular}{cccccccccc}\\
\hline 
$\sqrt{s_\mathrm{NN}}$ (TeV) &$y_\text{\,beam}$& $N_{1,2}$ & $N_\mathrm{gg}$ & $y_\mathrm{peak}$ & $\langle y_{1,2} \rangle$ & $\gamma_{1,2}$ & $\gamma_\mathrm{gg}$ & $\chi^2$ & $\chi^2 / \mathrm{ndf} $\\ 
\hline 
2.76 &$\pm 7.987$& 2700 & 12000 &$\pm 3.88$& $\pm 0.56$& 1000 & 115 & 5.89 & 0.16\\

5.02 &$\pm 8.586$& 2800 & 15800 & $\pm 4.43$& $\pm 0.61$& 2000 & 205 & 7.50 & 0.26 \\
\hline 
\label{tab3}
\end{tabular} 
\end{center}
\end{table}
\section{Conclusion}
We have investigated charged-hadron pseudorapidity distributions in central PbPb collisions within a three-source relativistic diffusion model with nonlinear drift, which ensures the 
correct Maxwell-J\"uttner equilibrium distribution. Our analysis indicates that the phenomenon of limiting-fragmentation scaling can be expected to hold at RHIC and LHC energies, spanning a factor of almost 260 in collision energy. This conclusion is in line with results from microscopic numerical models such as AMPT, but it disagrees with expectations from simple parametrizations of the rapidity distributions such as the difference of two Gaussians, and also with predictions from the thermal model. The latter does not explicitly treat the fragmentation sources, it refers only to particles produced from the hot fireball. 
In contrast, the fragmentation sources play an essential role in our approach.
It remains to be seen whether future upgrades of the detectors will make it possible to actually test the limiting-fragmentation conjecture experimentally at LHC energies.

\section*{Acknowledgment}

Discussions with multiparticle-dynamics group members Johannes H\"olck, Philipp Schulz (both ITP, Heidelberg) and Alessandro Simon (now Sophia University, Tokyo) are gratefully acknowledged.

\bibliographystyle{elsarticle-num}
\bibliography{kgw19.bib}
%

\end{document}